\documentclass[12pt]{article}
\usepackage{graphicx}
\begin{document} 

\title{Sznajd opinion dynamics with global and local neighbourhood}

\author{Christian Schulze\\
Institute for Theoretical Physics, Cologne University\\D-50923 K\"oln, Euroland}

\maketitle
\centerline{e-mail: ab127@uni-koeln.de}

\bigskip
Abstract:  
In this modification of the Sznajd consensus model on the square 
lattice, two people of arbitrary distance who agree in their
opinions convince their nearest neighbours of this opinion. 
Similarly to the mean field theory of Slanina and Lavicka, the
times needed to reach consensus are distributed exponentially
and are quite small. The width of the phase transition vanishes
reciprocally to the linear lattice dimension. Advertising has
effects independent of the system size. For more than two 
opinions, three opinions reach a consensus in roughly half of the
samples, and four only rarely and only for small lattices.


\bigskip
\bigskip

\begin{figure}[hbt]
\begin{center}
\includegraphics[angle=-90,scale=0.5]{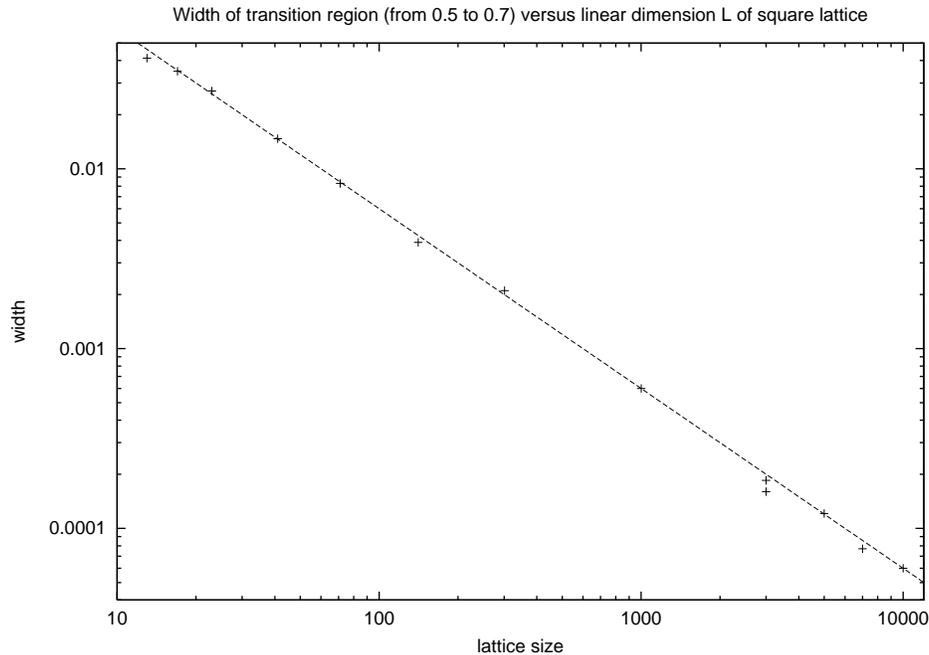}
\end{center}
\caption{ 
Width of the transition between consensus into opinion 2 to
consensus into opinion 1, based on 1000 lattices.
}
\end{figure}

\begin{figure}[hbt]
\begin{center}
\includegraphics[angle=-90,scale=0.5]{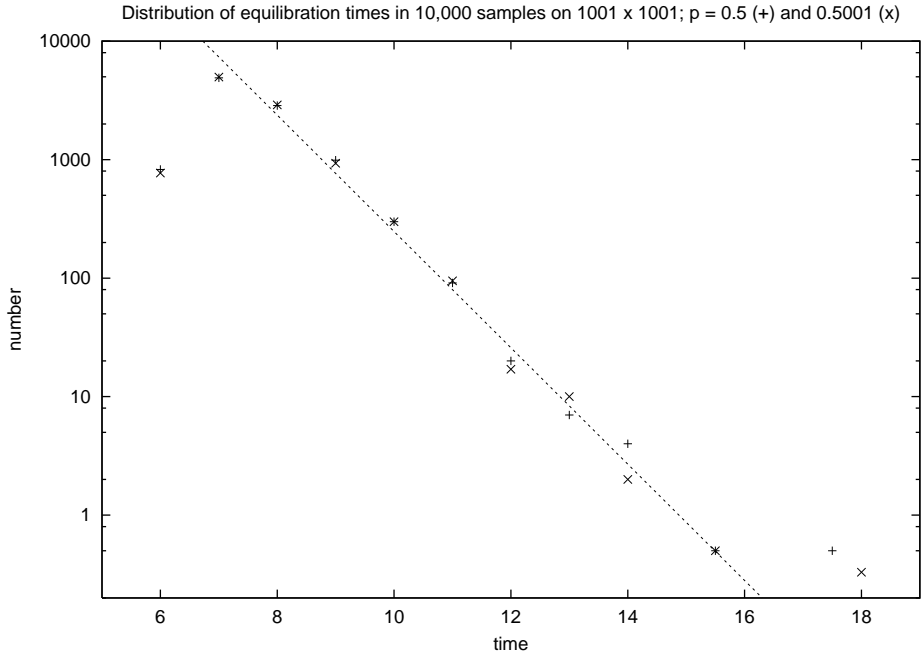}
\end{center}
\caption{ 
Histogram of the times needed to reach a consensus in 10,000
lattices of $1001 \times 1001$, showing exponential tail. ($Q=2$;
for $Q=3$ the behaviour is similar.)} 
\end{figure}

\begin{figure}[hbt]
\begin{center}
\includegraphics[angle=-90,scale=0.5]{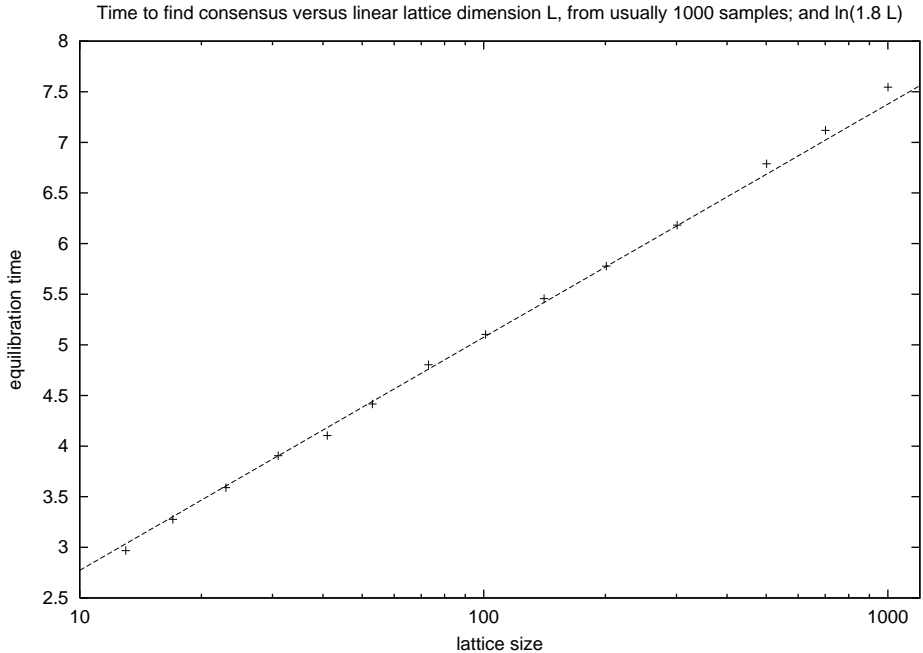}
\end{center}
\caption{ 
Variation of average time to reach consensus with the linear
lattice dimension $L$, from typically 1000 samples, showing
logarithmic increase as ln($L$) + const. ($Q=2$).
}
\end{figure}

\begin{figure}[hbt]
\begin{center}
\includegraphics[angle=-90,scale=0.5]{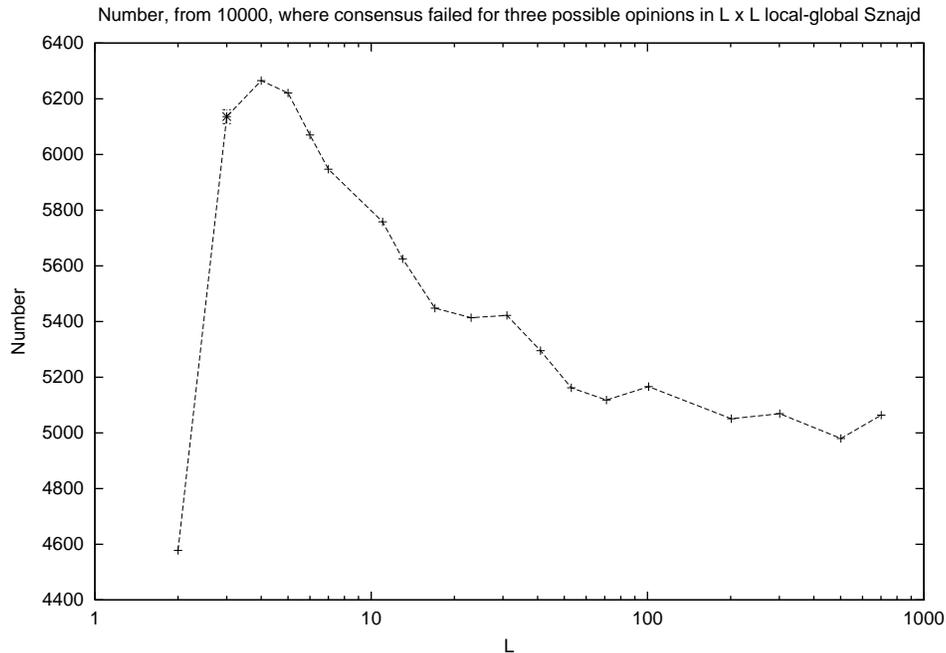}
\end{center}
\caption{ 
Failures, for 10,000 samples, to reach consensus with $Q = 3$
possible opinions. Failure means in most cases about equally many opinions
1 and 3, while consensus means everybody has opinion 2.
}
\end{figure}

\begin{figure}[hbt]
\begin{center}
\includegraphics[angle=-90,scale=0.5]{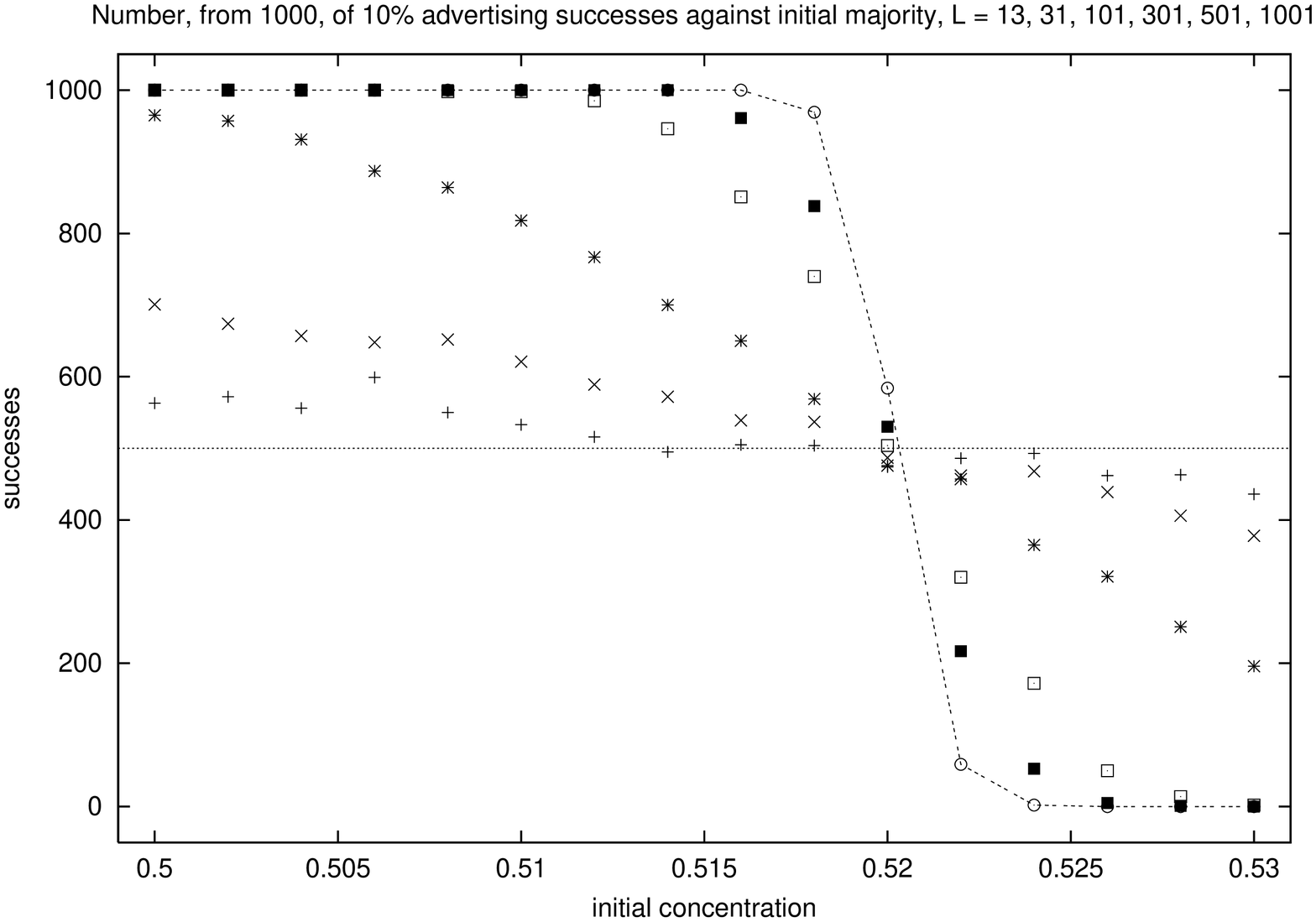}
\end{center}
\caption{ 
Advertising for opinion 2 wins in the left part and loses
in the right part against initial majority opinion 1, for $Q=2$. The 
larger the lattice is the sharper is the transition.
}
\end{figure}

Of the many recently simulated models of opinion dynamics 
\cite{axelrod,klemm,hegselmann,weisbuch,galam}, the Sznajd model
\cite{sznajd} was studied particularly often; see 
\cite{stauffer} for reviews and \cite{schulze,sornette}
for recent examples. In particular, Slanina and Lavicka
\cite{slanina} made a mean field theory and some simulations, 
where each pair of agents (spins) can be neighbours. This
infinite-range version agreed with many aspects of nearest 
neighbour interactions on square (or higher-dimensional)
lattices. The main difference was that for infinite range 
the time needed to reach a consensus was short and its
distribution decayed 
exponentially, while for nearest-neighbour square lattices
it was long and distributed in a more complicated way. The
present paper introduces an intermediate global-local model
with interactions both between arbitrarily selected agents 
and nearest neighbours, on the square lattice.

In this model, each of $N = L \times L$ 
lattice sites carries an agent which 
has one of $Q$ possible opinions. At each iteration we select
$N$ times two agents randomly; if and only if they share the same 
opinion, then each of the two agents convinces its four nearest 
neighbours on the lattice of the same opinion. Boundaries are 
neglected as usual \cite{stauffer}. If bounded confidence 
\cite{hegselmann,weisbuch} is used, only those neighbours are
convinced whose opinions differ by $\pm 1$ from the opinion of
the convincing pair; then opinion 2 can convince opinions 1 and 
3, but opinion 1 can convince only opinion 2, not opinion $Q$. 
If advertising \cite{schulzead,sznajd2} is used, then at each iteration 
with ten percent probability each agent independently adopts opinion 2,
independent of the previous opinion or that of the 
neighbours. Initially, the opinions are distributed randomly
and independently.

For the simple case $Q=2$ without advertising (bounded confidence
is meaningless at $Q = 2$) this new model shows the usual phase
transition: If the initial probability $p$ of opinions 1 is 
larger than 1/2, then at the end everybody shares this opinion 1;
if $p < 1/2$, everybody has opinion 2 at the end. In a finite 
lattice, the phase transition is rounded, and Fig.1 shows the 
width $W$ of the transition such that at initial concentration
$p = 0.5 + W$, of the 1000 samples about 700 show consensus for
opinion 1, and at $p = 0.5 - W$ only 300 of 1000 do so. Over 
three orders of magnitude in linear lattice dimension $L$ we see
$$ W \propto 1/L \quad .$$

We regard this simple power law as an indication that the model
belongs to the universality class of infinite range, even though
\cite{slanina} does not give such a finite-size scaling. For 
nearest-neighbour interactions, \cite{sousa} gave no clear
exponent. Another law $W \propto 1/L$ would follow from a 
simple majority rule for the initial opinions: On a finite 
lattice of $N$ sites the difference in the numbers of opinions
1 and 2 follows a binomial distribution about zero, which
for large $N$ approaches a Gaussian of width $\propto \sqrt N$ or
relative width $W  \propto 1/ \sqrt N = 1/L$. Thus the majority 
rule has a sharp phase transition for infinite systems, with
a finite width proportional to $1/L$ for finite lattices.

The reason why up to $L = 10,001$ could be simulated, far larger 
than any previous Sznajd lattice, is evident from Fig.2: 
Consensus is found after a few iterations, and the distribution
of the needed number $\tau$ of iterations decays exponentially
after a maximum at about 7. As a function of $L$, the average
time $<\tau>$ increases logarithmically, Fig.3. Both results
agree with the mean field theory of \cite{slanina}. 

If $Q = 3$ or 4, again a consensus is always found (not shown).
This changes if bounded confidence, as introduced above, 
restricts the convincing power. Then for $Q = 4$, a consensus 
is reached only rarely, and only for small lattices (not 
shown). For $Q = 3$, on the other hand, about half of the
10,000 simulated samples reach a consensus, Fig.4. For the
nearest-neighbour model, the border between consensus and
failure was between $Q = 3$ and 4, while now it is near 3.

With advertising and $Q = 2$, Fig.4 shows, in contrast to
the nearest-neighbour case \cite{schulzead,sznajd2}, a 
size-independent initial concentration of $p \simeq 0.52$
at which the two percent initial majority for opinion 1 
counteracts the ten percent advertising for opinion 2. 
(If advertising is doubled to 20 percent, the initial majority 
also doubles to four percent: $p \simeq 0.54$; not shown.)
In contrast, for nearest neighbour interaction 
\cite{sznajd2,schulzead}, advertising wins if the
system is large enough.

In summary, our global-local mixture of infinite-range 
interaction and
nearest-neighbour interaction is far superior numerically 
to nearest-neighbour interactions, and gives the short
relaxation times, distributed exponentially and increasing
as log($L$), of the mean field model \cite{slanina}.
 
\bigskip
Thanks are due to S. Havlin for suggesting to look for an upper critical
dimension, and D. Stauffer for help.



\begin{thebibliography}{99}

\bibitem{axelrod} R. Axelrod, {\it The Complexity of Cooperation}, Princeton
University Press 1997
\bibitem{klemm} Physica A
\bibitem{hegselmann} R. Hegselmann and M. Krause, 
\bibitem{weisbuch}
G. Weisbuch, G. Deffuant, F. Amblard, and J.-P. Nadal, 
Journal of Artificial Societies and Social Simulation 
5, No. 4, paper 1 (2002) (electronic only through jasss.soc.surrey.ac.uk).
Journal of Artificial Societies and Social Simulation 5, No.3,  paper 2 (2002)
\bibitem{galam} S. Galam, Physica A 333, 453 (2004).
\bibitem{sznajd} K. Sznajd-Weron and J. Sznajd, Int. J. Mod. Phys.
C 11, 1157 (2000).
\bibitem{stauffer} D. Stauffer, Journal of Artificial Societies 
and Social Simulation 5, No. 1, paper 4 (2000) and AIP Conference
Proceedings 690, 147 (2003).
\bibitem{schulze} D. Stauffer, A.O. Sousa and C. Schulze,
submitted to J. Artificial Societies and Social Simulation.
\bibitem{sornette} B. Roehner, D. Sornette and J.V. Andersen,
Int. J. Mod. Phys. C 15, issue 6 (2004).
\bibitem{slanina} F. Slanina and H. Lavicka, Eur. Phys. J. B, 35, 279 (2003).
\bibitem{schulzead} C. Schulze, Int. J. Mod. Phys. C 14, 95 (2003)
\bibitem{sznajd2} K. Sznajd-Weron and J. Sznajd, Physica A 324, 
437 (2003).
\bibitem{sousa} D. Stauffer, A.O. Sousa and S. Moss de Oliveira, Int. J. Mod. 
Phys. C 11, 1239 (2000).
\end{thebibliography}
\end{document}